\documentclass[10pt,a4paper]{article}
\usepackage[utf8]{inputenc}
\usepackage{amsmath}
\usepackage{amsfonts}
\usepackage{geometry}
\usepackage{amssymb}
\usepackage{graphicx}
\usepackage{authblk}

\title{Mass transport for Pollard waves}

\author[1]{Mateusz Kluczek}
\author[2]{Raphael Stuhlmeier}
\affil[1]{School of Mathematical Sciences, University College Cork, Cork, Ireland}
\affil[2]{Centre for Mathematical Sciences, University of Plymouth, Plymouth, UK}
\date{}

\begin{document}
\maketitle

\begin{abstract}
We provide an in-depth exploration of the mass-transport properties of Pollard's exact solution for a zonally-propagating surface water-wave in infinite depth. Without resorting to approximations we discuss the Eulerian mass transport of this fully nonlinear, Lagrangian solution. We show that it has many commonalities with the linear, Eulerian wave-theory, and also find Pollard-like solutions in the first and second order Lagrangian theory.
\end{abstract}

\maketitle
\section{Introduction}
In 1970 Pollard \cite{Pollard1970} introduced a modification of the famous Gerstner wave \cite{Gerstner1804} which provided an exact solution for the Euler equations with Coriolis forces. This was the first of many -- and in recent years increasingly complex -- exact solutions for geophysical waves (see \cite{Henry2017a,Ionescu-Kruse2017,Johnson2017} and references therein). Pollard's zonally propagating wave has since been studied by a number of authors -- its stability was investigated by Ionescu-Kruse \cite{Ionescu-Kruse2016}, who also considered the restriction to the equatorial $f$-plane \cite{Ionescu-Kruse2015}, and it was recently extended to internal waves by Kluczek \cite{Kluczek2019,Kluczek2018}. A thoroughgoing investigation of Pollard's solution, including the effects of an underlying current, was performed by Constantin \& Monismith \cite{ConstantinMonismith2017}. Rodriguez-Sanjurjo \cite{Rodriguez-Sanjurjo2018a} showed rigorously that a Pollard-like solution is globally dynamically possible. While the solution is rotational and thus the associated flow cannot be generated by conservative forces, some observations indicate that such waves cannot be ruled out in wave tank experiments \cite{Monismith2007,Weber2011}. Further discussion of the oceanographic relevance of these solutions may be found in Boyd \cite{Boyd2018}.

Pollard's seminal work explored the mass transport of surface waves with rotation, for the exact Lagrangian solution, as well as for Stokes' solution. The present paper employs new techniques and numerical tools to shed further light on the matter, treating a number of issues exactly for the first time. While the Lagrangian mass transport for Pollard's wave is immediate, we are also able to compute the Eulerian mass transport exactly using the Lambert W-function \cite{Valluri2000}. Furthermore, in light of the surprising commonalities that have been found between trochoidal, exact solutions and linear theory (e.g.\ \cite{Stuhlmeierc,Kinsman1984}), we explore the linearized problem with rotation in both Eulerian and Lagrangian coordinates.

In what follows, we first present the governing equations in Lagrangian and Eulerian coordinates in Section \ref{sec: Governing equations}. We introduce Pollard's solution, and discuss the Eulerian and Lagrangian mean velocities associated with it in Section \ref{sec: Pollard solution}. In Section \ref{sec: Linear Eulerian} we compare Pollard's solution with the solution to the linearized problem in Eulerian coordinates, calculating the associated Eulerian and Lagrangian mean velocities. In Section \ref{sec: Linear Lagrangian} we look at the corresponding theory for the linearized problem in Lagrangian coordinates. Finally, we present some concluding remarks in Section \ref{sec: Conclusions}.

\section{Governing equations}
\label{sec: Governing equations}

The governing equations for water waves in the presence of Coriolis forces may be found in most textbooks on geophysical fluid dynamics, e.g.\ Pedlosky \cite{Pedlosky1982}.  The local reference frame with origin at the Earth's surface, and rotating together with the Earth is presented in Cartesian coordinates. Therefore, the $(x,y,z)$ coordinates represent the direction of the latitude, longitude and local vertical, respectively (sometimes called zonal, meridional, and vertical). The respective components of the velocity field are $\mathbf{u}=(u,v,w),$ and the governing equations take the form
\begin{equation}\label{eq:Governing equation}
  \left\lbrace
    \begin{aligned}
      &u_t+uu_x+vu_y+wu_z+\hat{f}w-fv=-\dfrac{1}{\rho}P_x,\\
      &v_t+uv_x+vv_y+wv_z+fu=-\dfrac{1}{\rho}P_y,\\
      &w_t+uw_x+vw_y+ww_z-\hat{f}u=-\dfrac{1}{\rho}P_z-g,
    \end{aligned}
  \right.
\end{equation}
where the Coriolis parameters are given by
\begin{equation*}\label{eq:Coriolis parameters}
  f=2\Omega\sin\phi, \qquad \hat{f}=2\Omega\cos\phi.
\end{equation*}
Here $\Omega$ is the constant rotational speed of the Earth, $\Omega=7.29\times10^{-5}$rad$\cdot s^{-1}$ around the polar axis towards the east and $\rho$ denotes the density of fluid. An assumption of constant density yields the incompressibility condition
\begin{equation*}\label{eq:Incompressibility}
  u_x+v_y+w_z=0.
\end{equation*}
Finally, we must impose conditions on the sea-surface, given by $z=\eta(x,y,t):$
\begin{equation}\label{eq:Boundary conditions}
  \begin{aligned}
    &P=P_{atm} &\text{ on the free surface } z=\eta(x,y,t),\\
    &w=\eta_t+u\eta_x+v\eta_y &\text{ on the free surface } z=\eta(x,y,t),
  \end{aligned}
\end{equation}
where $P_{atm}$ denotes the constant atmospheric pressure.
For the purpose of describing waves in the open ocean, the deep-water assumption of vanishing velocities with depth supplements the above.

In the Lagrangian description, the same equations take the form
\begin{align} \label{eq:Lagrangian GE 1}
  &(x_{tt} - fy_t + \hat{f}z_t)x_q + (y_{tt}+fx_t)y_q + (z_{tt} - \hat{f}x_t)z_q + P_q/\rho + g z_q = 0,\\ \label{eq:Lagrangian GE 2}
  &(x_{tt} - fy_t + \hat{f}z_t)x_r + (y_{tt}+fx_t)y_r + (z_{tt} - \hat{f}x_t)z_r + P_r/\rho + g z_r = 0,\\ \label{eq:Lagrangian GE 3}
  &(x_{tt} - fy_t + \hat{f}z_t)x_s + (y_{tt}+fx_t)y_s + (z_{tt} - \hat{f}x_t)z_s + P_s/\rho + g z_s = 0,
\end{align}
with the incompressibility condition that the Jacobi determinant be time-independent, i.e.\
\begin{equation*}\label{eq: Lagrangian incompr}
  \frac{d}{dt} \vrule \frac{\partial(x,y,z)}{\partial(q,r,s)} \vrule = 0,
\end{equation*}
for particle labels $q,\, r,$ and $s$ and positions $x(q,r,s,t), \, y(q,r,s,t)$ and $z(q,r,s,t),$ respectively.

\section{Pollard's exact solution}
\label{sec: Pollard solution}

In terms of the Lagrangian labels $q,\, r,$ and $s,$ Pollard \cite{Pollard1970} gave an explicit solution in the following form:
\begin{equation}\label{eq:Pollard explicit solution}
  \left\lbrace
    \begin{aligned}
      &x=q-be^{ms}\sin\theta,\\
      &y=r-de^{ms}\cos\theta,\\
      &z=s+ae^{ms}\cos\theta.
    \end{aligned}
  \right.
\end{equation}
where $\theta=k(q-ct),$ and
\begin{equation*}\label{eq:Labelling variables}
  (q,r,s)\in\mathbb{R} \times(-r_0,r_0)\times\left(-\infty,s_0(r)\right).
\end{equation*}
Here the Lagrangian labelling variable $r\in(-r_0,r_0)$ since we resolve the solution around a fixed latitude $\phi,$ so that $f$ and $\hat{f}$ are constant. The parameter $s_0(r)$ represents the free surface, $k$ the wavenumber, $c$ the wave speed, $t$ the time and $a$ the fixed amplitude.

\begin{figure}
  \centering
  \includegraphics[scale=0.6]{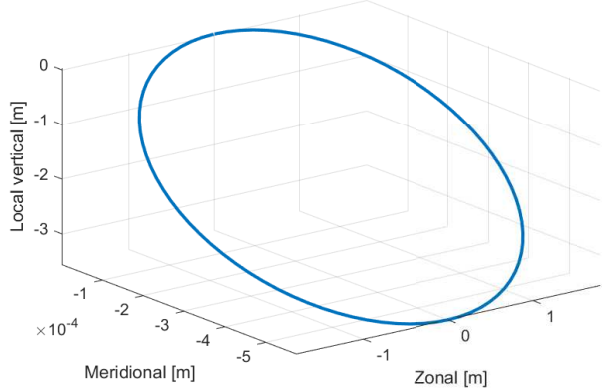}
  \caption{Closed particle paths of Pollard's wave. Here $k=0.0628 \text{m}^{-1},\, \phi=60^\circ$ and $a=2$ m. }
\label{fig:Particle_trajectory}
\end{figure}

\begin{figure}
  \centering
  \includegraphics[width=0.8\textwidth]{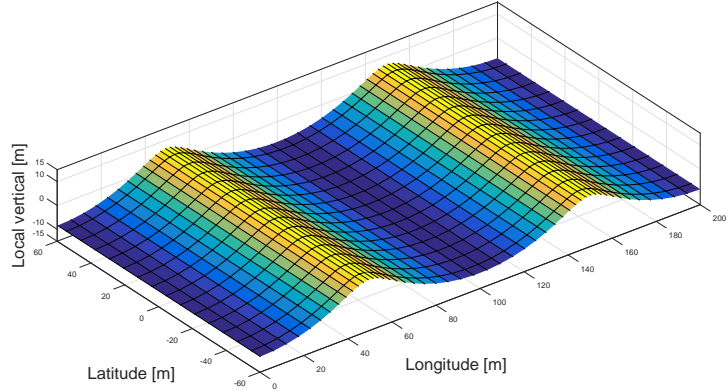}
  \caption{The profile of the wave with the parameters $k=6.28\times10^{-2}$ $m^{-1}$, $c=12.5$ m$s^{-1}$, $a=8$ m propagating at $\phi=60^\circ$.}\label{fig:The profile_of_free_surface}
\end{figure}

\begin{figure}
  \centering
  \includegraphics[width=0.7\textwidth]{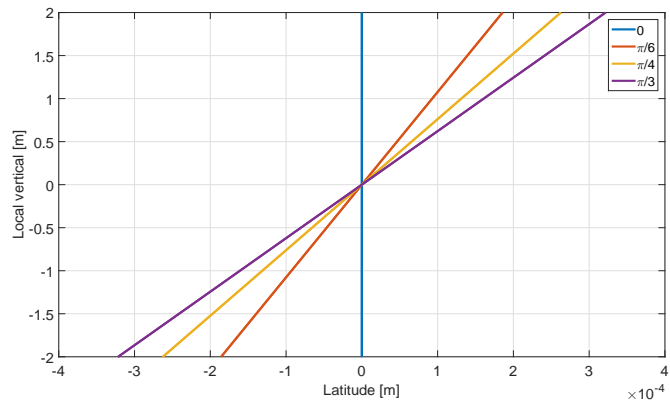}
  \caption{The tilt of particle orbits at latitudes 0$^\circ$, 30$^\circ$, 45$^\circ$ and 60$^\circ$ South. The tilt of particles increases as we move away from the equator towards the poles. The parameter of the amplitude is fixed to the local vertical, therefore the vertical cross section of the wave does not change with increasing latitude.}\label{fig:The tilt of particles}
\end{figure}

Using the governing equations, we find the following relations between the parameters occurring in \eqref{eq:Pollard explicit solution}:
\begin{align}
  & m = \frac{ck^2}{\gamma} \label{eq: m}\\
  & b = \frac{ack}{\gamma} \label{eq: b}\\
  & d = -\frac{af}{\gamma} \label{eq: d}
\end{align}
with $\gamma = \sqrt{c^2k^2 - f^2}.$ The pressure is then given explicitly by

\begin{equation} \label{eq: Pollard pressure}
  P = \rho \frac{1}{2}e^{2ms} a^2 c^2 k^2 \left( 1+ \frac{ \hat{f}}{\gamma} \right) - \rho g s + P_{atm}.
\end{equation}
The dispersion relation is given by

\begin{equation}\label{eq:Dispersion relation}
  c^2\left(c^2k^2-f^2\right)=\left(g-\hat{f}c\right)^2.
\end{equation}
For vanishing rotation $\Omega =0$, Pollard's solution \eqref{eq:Pollard explicit solution} reduces to the famous Gerstner wave, with the addition of a parameter $a.$ Otherwise it is a rotationally modified, zonally propagating wave with trochoidal free surface (see Figure \ref{fig:The profile_of_free_surface}), whose particle trajectories are closed circles inclined at an angle to the local vertical (see Figure  \ref{fig:Particle_trajectory}). The inclination of the particle paths increases with distance from the equator (see Figure \ref{fig:The tilt of particles}).

\subsection{Eulerian mean velocities}
\label{ssec: Pollard - Eulerian velocities}

Calculation of Eulerian mean velocities requires fixing a point $(x_0,y_0,z_0)$ in the fluid domain, and integrating the (Eulerian) velocity field $u(x_0,y_0,z_0,t)$ over one wave period. For Pollard's exact solution, this means inverting \eqref{eq:Pollard explicit solution}, so that $u(q,r,s,t)=\dot{x}(q,r,s,t)$ (for fixed labels $q,r,s$ and varying positions $x,y,z$) can be recast as $u(x,y,z,t)$ (for fixed $x,y,z$ and varying labels $q,r,s$). Such a procedure cannot be carried out analytically -- however we shall see that the Eulerian mean velocity depends only on depth $z_0,$ and will compute it numerically.

Fixing a depth $z_0 = s + a e^{ms} \cos(\theta)$ beneath the wave trough, it is clear that we can express implicitly $s=S(z_0,\theta).$ In fact, denoting by $W$ the Lambert W-function, $S=z_0-W(a m \cos(\theta) e^{m z_0})/m$ . Thus, noting that $u$ is independent of $y,$

\begin{equation}\label{eq:Calculation of mean Eulerian velocity zonal}
  \begin{aligned}
    c+\left<u\right>_E(z_0) & =\dfrac{1}{T}\int_{0}^{T}[c+u(x-ct,z_0)]dt=\\
    &=\dfrac{1}{L}\int_{0}^{L}[c+u(x-ct,z_0)]dx=\\
    &=\dfrac{1}{L}\int_{0}^{L}[c+u(q-ct,z_0)] \dfrac{\partial x(q,S)}{\partial q}dq=\\
    &=\dfrac{1}{L}\int_0^L(c+kcbe^{ms}\cos\theta)\dfrac{1-a^2m^2e^{2ms}}{1+mae^{ms}\cos\theta}dq=\\
    &=c-\frac{m^2a^2c}{L}\int_{0}^{L}e^{2mS}dq.
  \end{aligned}
\end{equation}
Therefore, the mean Eulerian velocity in the longitudinal direction is

\begin{equation}\label{eq:Zonal mean Eulerian velocity}
  \left<u\right>_E(z_0) = -\frac{m^2a^2c}{L}\int_{0}^{L}e^{2mS(z_0,\theta)}dq.
\end{equation}
which depends only on $z_0$ since $S$ is periodic in $\theta = q-ct.$ This exact expression may be compared with the leading order approximation given in eq.\ (25) of Pollard \cite{Pollard1970}.

\begin{figure}
  \centering
  \includegraphics[width=0.7\textwidth]{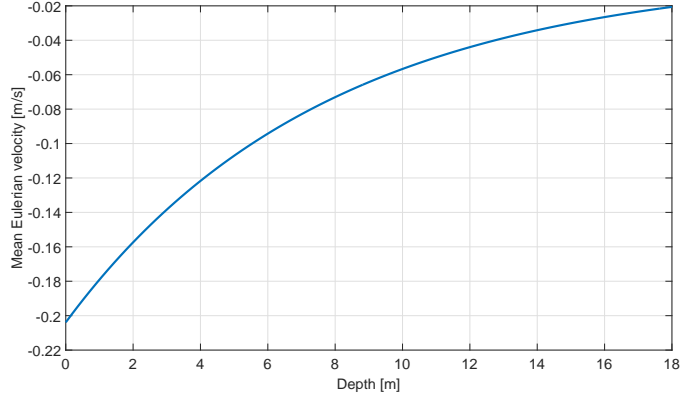}
  \caption{The mean Eulerian velocity depicted for different depths. The mean velocity is evaluated at latitude $\phi=60^\circ$. The respective parameters of the wave are $k=6.28\times10^{-2}$ $m^{-1}$, $c=12.5$ m$s^{-1}$, $a=2$ m with period $T=8$ s.}\label{fig:MEV_wrt_depth}
\end{figure}

\begin{figure}
  \centering
  \includegraphics[width=0.7\textwidth]{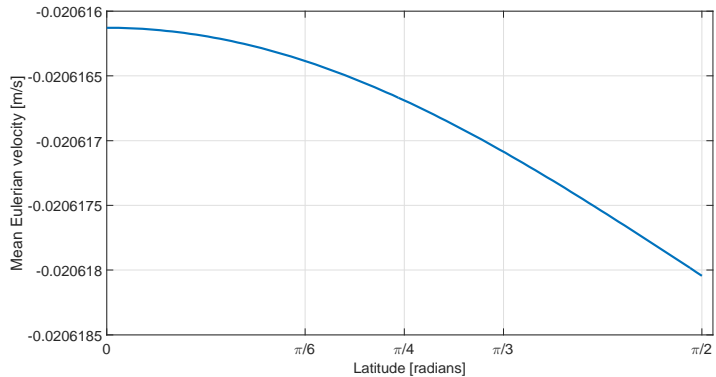}
  \caption{The mean Eulerian velocity for a fixed point at the depth $z=-18$ m. The plot presents the mean velocity evaluated with respect to increasing latitude. The wavenumber is $k=6.28\times10^{-2}$ $m^{-1}$, the wave speed is $c=12.5$ m$s^{-1}$, and the amplitude of the wave at the surface is $a=2$ m with period $T=8$ s.}\label{fig:MEV_wrt_latitude}
\end{figure}
The Eulerian mean velocity is negative, and largest at the equator, where $\phi=0,$ and falls off towards the poles. This may be seen in Figure \ref{fig:MEV_wrt_latitude}, where \eqref{eq:Zonal mean Eulerian velocity} has been evaluated numerically for different values of $\phi.$ It is readily verified that

\begin{equation*}\label{eq: Derivative of mean eulerian wrt depth}
  \frac{\partial}{\partial z_0}\langle u \rangle_E(z_0)= -\frac{m^2a^2c}{L}\int_{0}^{L}S_{z0} e^{2mS(z_0,\theta)}dq >0,
\end{equation*}
so that $\langle u \rangle_E$ decreases with increasing depth. Numerical evaluation of \eqref{eq:Zonal mean Eulerian velocity} shows this decrease in mean Eulerian velocity, as depicted in Figure \ref{fig:MEV_wrt_depth}. There are no Eulerian mean velocities in the vertical or meridional direction, as $\langle v \rangle_E$ and $\langle w \rangle_E$ lead to an integral akin to \eqref{eq:Calculation of mean Eulerian velocity zonal}, albeit with an odd, L-periodic integrand.

\subsection{Lagrangian mean velocities}
\label{ssec: Pollard - Lagrangian velocities}

The Lagrangian mean velocities may be calculated from \eqref{eq:Pollard explicit solution} by taking the integral averaged over a wave period. For the longitudinal Lagrangian mean velocity

\begin{equation} \label{eq: Pollard Lagrangian mean u}
  \langle u \rangle_L = \frac{1}{T}\int^T \dot{x}(q,r,s,t) \text{ d}t,
\end{equation}
with analogous expressions for the meridional and vertical mean velocities. Differentiating \eqref{eq:Pollard explicit solution} w.r.t.\ $t$ and inserting into \eqref{eq: Pollard Lagrangian mean u} we see that all Lagrangian mean velocities vanish, reflecting the fact that the particle trajectories are closed (see Figure \ref{fig:Particle_trajectory}) and there is no net wave transport.

\section{Linear Eulerian theory for geophysical surface waves}
\label{sec: Linear Eulerian}

The Eulerian equations \eqref{eq:Governing equation}--\eqref{eq:Boundary conditions} may be linearised by assuming small steepness of the surface waves. The linear equations for waves travelling in the zonal direction, analogous to Pollard's waves, were given by Constantin \& Monismith \cite{ConstantinMonismith2017}

\begin{align*}
  &u_t + \hat{f}w - fv = -P_x,\\
  &v_t + fu = -P_y, \\
  &w_t - \hat{f}u = -P_z - g,\\
  &u_x +v_y + w_z = 0, \\
  &w = \eta_t \text{ on } z = 0,\\
  &u_t + \hat{f}w - fv + g\eta_x = 0 \text{ on } z = 0,
\end{align*}\label{eq: Linearised GE}
with solution

\begin{align}
  &\eta = a_0 \cos(\theta),\nonumber\\
  &u = mca_0 e^{mz} \cos(\theta), \label{eq:lin_euler_u}\\
  &v = \frac{fma_0}{k}e^{mz} \sin(\theta), \label{eq:lin_euler_v}\\
  &w = kca_0 e^{mz} \sin(\theta), \label{eq:lin_euler_w}
\end{align}\label{eq: Linearised Eulerian sol}
where $\theta = k(x-ct),$ and $m$ and $c$ are given in \eqref{eq: m} and \eqref{eq:Dispersion relation}, respectively.

\subsection{Eulerian mean velocities}
\label{ssec: Linear Eulerian - Eulerian velocities}

The Eulerian mass transport is defined as the average

\begin{equation*}\label{eq: mean linear Eulerian vel}
  \langle u \rangle_E  = \frac{1}{T} \int_0^T \mathbf{u} \, dt,
\end{equation*}
which is readily seen to vanish below the trough line for all components of the velocity field \eqref{eq:lin_euler_u}--\eqref{eq:lin_euler_w}. By Taylor expansion about the free surface level (see \cite[p.\ 285ff]{Dean1991}), the zonal Eulerian mean velocity at the surface $\langle u(x,z,\eta) \rangle_E = a_0^2 m^2 c/2,$ while the meridional and vertical Eulerian mean velocities vanish. This second order effect arising from the linear theory was not considered in Pollard.

\subsection{Lagrangian mean velocities}
\label{ssec: Linear Eulerian - Lagrangian velocities}

To elucidate the Lagrangian mean velocities $\langle u \rangle_L$ we must consider the particle trajectories.
These satisfy the system of coupled, nonlinear ODEs
\begin{align}\label{eq: Particle Trajectory ODEs1}
  \dot{x} = u,\,  \dot{y} = v,\, \dot{z} = w,
\end{align}
for $u, \, v, \, w$ given in \eqref{eq:lin_euler_u}--\eqref{eq:lin_euler_w} above. By suitable rotation of the coordinates, akin to that adopted for edge waves \cite{Stuhlmeier2015}, it can be shown that the Lagrangian particle paths in this linear, Eulerian theory are non-closed, with  a forward drift over the wave period, corresponding to a positive Lagrangian mean velocity in the direction of wave propagation.

\begin{figure}
  \centering
  \includegraphics[scale=0.6]{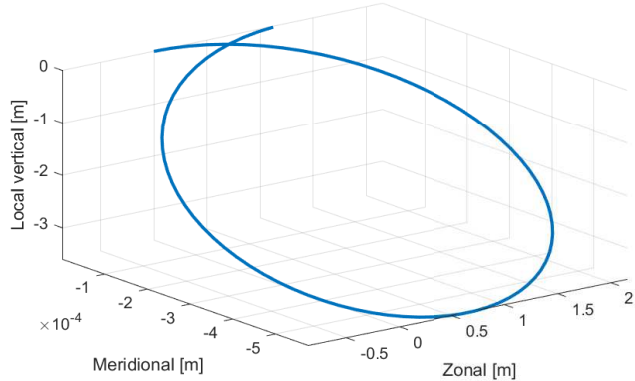}
  \caption{Example particle paths, obtained by numerical integration of the linear particle trajectory ODEs over one period. Here $k=0.0628 \text{m}^{-1},\, \phi=60^\circ$ and $a=2$ m. The initial position is chosen as $(x_0,y_0,z_0)=(0,0,0).$ The drift over one period is 1.1 m in the direction of propagation.}
  \label{fig: Particle Paths}
\end{figure}

We switch to the rotated coordinates

\begin{equation*}\label{eq: Rotation}
  \begin{pmatrix}
    x'\\
    y'\\
    z'
  \end{pmatrix}
  =
  \begin{pmatrix}
    1 & 0 & 0 \\
    0 & \cos \alpha & \sin \alpha \\
    0 & -\sin \alpha & \cos \alpha
  \end{pmatrix}
  \begin{pmatrix}
    x\\
    y\\
    z
  \end{pmatrix}
\end{equation*}
and let $\tan \alpha = d/a = -f/\gamma,$ see \eqref{eq: d}. In this rotated frame, the particle trajectory ODEs have the form
\begin{align*}
  &\frac{d x'}{dt}=mca_0e^{m(y'\sin \alpha + z' \cos \alpha)} \cos(\theta),\\
  &\frac{d y'}{dt}=0,\\
  &\frac{d z'}{dt}=mca_0e^{m(y'\sin \alpha + z' \cos \alpha)}\sin(\theta).
\end{align*}\label{eq: Particle trajectory ODEs}
Thus the particle motion is two-dimensional, confined to planes inclined at an angle $\alpha$ to the vertical (see \cite{ConstantinMonismith2017}), and using the results of \cite{Constantin2008e} the particle paths are found to be non-closed, as is easily seen using numerical integration (see Figure \ref{fig: Particle Paths}). Prior work, including Pollard's \cite[Eq.\ (24)]{Pollard1970} built on integrating the particle trajectory ODEs \eqref{eq: Particle Trajectory ODEs1} for a small displacement about the initial position $(x_0,y_0,z_0)$ only (see, e.g.\, \cite[Sec.\ 4.2.2]{Dean1991}), thereby circumventing the Lagrangian mass transport contained in the first-order theory.

While a fully rigorous, quantitative theory is out of reach, by numerical integration it may be established easily that the first-order drift obtained by integrating \eqref{eq: Particle Trajectory ODEs1} is of roughly the same magnitude as the Stokes' drift.

\section{Linear Lagrangian theory for geophysical surface waves}
\label{sec: Linear Lagrangian}

Following Pierson \cite{2011}, one can obtain a solution to the linearized form of the Lagrangian governing equations \eqref{eq:Lagrangian GE 1}--\eqref{eq:Lagrangian GE 3}. Linearizing about the hydrostatic solution  ($x=q,\, y=r,\, z=s$) yields

\begin{align}\label{eq:Lagrangian GE 1 lin}
  &x_{tt} - fy_t + \hat{f}z_t + P_q/\rho + g z_q = 0,\\ \label{eq:Lagrangian GE 2 lin}
  &y_{tt}+fx_t + P_r/\rho + g z_r = 0,\\ \label{eq:Lagrangian GE 3 lin}
  &z_{tt} - \hat{f}x_t + P_s/\rho + g z_s = 0,
\end{align}
and the linear incompressibility condition

\begin{equation}\label{eq: Lagrangian incomp lin}
  x_{qt}+y_{rt}+z_{st} = 0.
\end{equation}
It is remarkable that a Pollard-like wave also solves \eqref{eq:Lagrangian GE 1 lin}--\eqref{eq: Lagrangian incomp lin}. Making an ansatz of the form \eqref{eq:Pollard explicit solution} leads to a set of algebraic equations, and the relations

\begin{equation*}\label{eq: Parameters for linearised lagrangian GE}
  m= \frac{k^2(\hat{f}c - g)}{f^2-c^2k^2}, \quad b = \frac{am}{k}, \quad d=\frac{-amf}{ck^2},
\end{equation*}
cf.\ \eqref{eq: m}--\eqref{eq: d}. Subsequently the dispersion relation \eqref{eq:Dispersion relation} is recovered, and it can be seen that the particle paths are circular, owing to $a^2+d^2=b^2.$ The vertical decay scale, controlled by $m$, for the linear solution is distinct from the full, nonlinear solution, and the pressure is hydrostatic $P=P_{atm} - \rho g s.$ The exact pressure \eqref{eq: Pollard pressure} of Pollard's exact solution is subsequently recovered as a second order correction.

\section{Discussion}
\label{sec: Conclusions}
Exact, trochoidal solutions have much in common with the periodic wave trains found in linear Eulerian theory. In both cases we find a fluid velocity field exhibiting exponential decay with depth together with sinusoidal motion in the direction of propagation. Despite the fact that linear theory posits small wave slope, while the exact theory is valid up to the theoretical steepest (cusped) wave, we find the same dispersion relation in both cases. These commonalities have been discussed elsewhere for Gerstner waves and trochoidal edge waves, and we find them again for Pollard's rotationally modified trochoidal wave.

The exact, trochoidal theory relies on tracking particle labels, and is based on the Lagrangian form of the governing equations. This gives rise to the two major discrepancies between linear Eulerian waves and the trochoidal theory: vorticity and mass transport. We have undertaken a study of the mass transport of Pollard's solution, deriving in \eqref{eq:Zonal mean Eulerian velocity} an expression for the Eulerian mean velocity in terms of the Lambert W-function. This depends only on depth and latitude, and can easily be evaluated numerically. The Lagrangian mean velocities of Pollard's solution are zero due to the closed particle paths.

The mass transport for the linear, Eulerian solution was also investigated, where Coriolis forces have the effect of slightly modifying the classical, sinusoidal wave. Nevertheless, we highlight the impact of recent results showing that the particle trajectories are not closed even for linear theory, and demonstrate the impact of this on the Lagrangian mass transport for linear, Eulerian waves in a rotating frame. By a suitable rotation of the coordinate frame, we see that the motion in these waves is strictly 2D, and particles undergo a drift in the direction of wave propagation. This first order--drift is comparable in magnitude to the second--order Stokes' drift. We also demonstrate that the Pollard wave solves the linearized Lagrangian equations with hydrostatic pressure, and also satisfies the second-order Lagrangian equations.


\subsection*{Acknowledgment}
Mateusz Kluczek acknowledges gratefully the support of the Science Foundation Ireland (SFI) research grant 13/CDA/2117.

\end{document}